\documentclass[fleqn,usenatbib]{mnras}

\usepackage[T1]{fontenc}
\usepackage{ae,aecompl}

\usepackage{graphicx}	
\usepackage{amsmath}	
\usepackage{amssymb}	
\usepackage{pdflscape}
\usepackage{times,txfonts}
\usepackage{color}
\usepackage{hyperref}	

\definecolor{linkcolor}{rgb}{0,0,0.25}
\hypersetup{
  colorlinks=true,        
  linkcolor=linkcolor,    
  citecolor=linkcolor,    
  filecolor=linkcolor,    
  urlcolor=linkcolor      
}

\usepackage{multirow}
\usepackage{rotating}
\usepackage{lscape}
\usepackage{tablefootnote}
\usepackage{hyperref}
\usepackage{listings}

\usepackage[colorinlistoftodos]{todonotes}

\newcommand{\gaia}{\emph{Gaia}}

\title[The GD-1 Stream Progenitor]{Searching for the GD-1 Stream Progenitor in \gaia\ DR2 with Direct N-body Simulations}
\author[Webb \& Bovy]{Jeremy J. Webb\thanks{E-mail: webb@astro.utoronto.ca (JW)} \& Jo Bovy
 \\
Department of Astronomy and Astrophysics, University of Toronto, 50 St. George Street, Toronto, ON, M5S 3H4, Canada
}

\begin{document}

\pagerange{\pageref{firstpage}--\pageref{lastpage}} \pubyear{2018}

\maketitle

\label{firstpage}

\begin{abstract}

We perform a large suite of direct N-body simulations aimed at revealing the location of the progenitor, or its remnant, of the GD-1 stream. Data from \gaia\ DR2 reveals the GD-1 stream extends over $\approx 100^\circ$, allowing us to determine the stream's leading and trailing ends. Our models suggest the length of the stream is consistent with a dynamical age of between 2-3 Gyr and the exact length, width and location of the GD-1 stream correspond to the stream's progenitor being located between $-30^\circ < \phi_{1,\mathrm{pro}} < -45^{\circ}$ in the standard GD-1 coordinate system. The model stream density profiles reveal that intact progenitors leave a strong over-density, recently-dissolved progenitors appear as gaps in the stream as escaped stars continue to move away from the remnant progenitor's location, and long-dissolved progenitors leave no observational signature on the remaining stream. Comparing our models to the GD-1 stream yields two possible scenarios for its progenitor's history: a) the progenitor reached dissolution approximately 500 Myr ago during the cluster's previous perigalactic pass and is both located at and responsible for the observed gap at $\phi_1=-40^{\circ}$ or b) the progenitor reached dissolution over 2.5 Gyr ago, the fully-dissolved remnant is at $-30^\circ < \phi_1 < -45^{\circ}$, and an observational signature of its location no longer exists. That the dissolved progenitor is in the range $-30^\circ < \phi_1 < -45^{\circ}$ implies that density fluctuations outside of this range, e.g., a deep gap at $\phi_1 \approx -20^\circ$, are likely produced by compact baryonic or dark-matter perturbers.
\end{abstract}


\section{Introduction} \label{intro}
Star clusters form an important component of the stellar distribution of galaxies. As they orbit within their host galaxy's gravitational potential, tidal forces strip away the most weakly-bound stars and escaped stars continue along the cluster's orbital path with orbital energies that are either slightly lower or slightly higher than their progenitor cluster. As stars are stripped over the history of the cluster, a stellar stream forms that is kinematically cold. Many such cold stellar streams are observed in the Milky Way (see \citet{grillmair16} for a recent review). The fact that interstellar forces within a stream are near negligible means that the stream's track effectively traces out the stream's orbit, making them highly useful in mapping out the gravitational potential of galaxies like the Milky Way \citep[e.g.,][]{bovy16_mwmap}. 

With some stellar streams spanning tens of degrees across the sky, which can translate into having lengths of several kiloparsecs, streams have a high-cross section for interacting with the various forms of substructure in the Galaxy. This feature has led to streams being one of the most often used tools in the search for evidence of the existence of dark matter sub-halos, which the Cold Dark Matter framework predicts should make up between $0.1\,\%$ and $10\,\%$ of the dark matter density in the Galaxy, depending on the stream's Galactocentric distance \citep{diemand08,springel08}. Theoretical models predict that a sub-halo passing through a stream will create a kinematic perturbation along the stream at the sub-halo's crossing point. If the interaction has occurred recently enough, then the local velocity dispersion of stars should be higher than those along the stream. Over time, a visible under-density in stars along the stream will form as stars move away from where the sub-halo crossed the stream \citep[e.g.,][]{johnston02, ibata02, carlberg09, erkal16,sanders16,bovy17,carlberg17}. Hence stream gaps can be used as a tracer for the properties of dark matter substructure in the Milky Way \citep{yoon11, carlberg12, erkal15a, erkal15b, bovy16, carlberg16,bovy17,Banik18b}.

Unfortunately, dark matter sub-halo interactions are not the only perturbing objects that can create a gap along a stellar stream. Interactions with giant molecular clouds, the Galactic bar, spiral arms, perigalactic passes, and disc passages have been shown to also affect the properties of stellar streams \citep[e.g.,][]{Amorisco16a,pearson17,banik18}. The progenitor cluster from which the stream formed will also leave a signature along the stream's track in the form of either an over-density---if the cluster has yet to fully dissolve---or an under-density---as stars continue to move away from where the cluster would be had it not dissolved. Hence it is important to identify the location of a stream's progenitor and the effects of other perturbers before using a stream to study dark matter substructure.

The GD-1 stream, first discovered by \citet{grillmair06}, is perhaps one of the most heavily studied stellar streams due to the fact that it spans approximately $80^{\circ}$ across the sky and is located close to the Sun ($~10$ kpc) while remaining quite narrow ($~2^{\circ}$ or $70$ (pc)), making it easily identifiable in observational surveys \citep{carlberg13, pricewhelan18}. Along the stream's track there exists a series of gaps and over-densities \citep{koposov10, carlberg13,pricewhelan18,deboer18} that may be due to interactions with dark-matter substructure. Recently it was suggested that an off-stream spur of stars also exists, indicating a possible interaction between the stream and a dark-matter sub-halo \citep{bonaca18}. Unfortunately, studies that use the properties of the GD-1 stream to place constraints on the dark matter substructure fraction of the Milky Way are limited by the fact that the location of GD-1's progenitor cluster or progenitor remnant remains unknown.

GD-1 is traditionally studied in a celestial coordinate system $(\phi_1,\phi_2)$ in which the stream is approximately at $\phi_2 =0$ that was introduced by \citet{koposov10}. In this coordinate system, the leading candidates for the location GD-1's progenitor are $\phi_1 = -13^{\circ}$, $-20^{\circ}$, and $-45^{\circ}$. Making use of \gaia\ DR2 astrometry and Pan-STARRS photometry, \citet{pricewhelan18} finds a significant over-density of stars at $\phi_1 = -13^{\circ}$ that could correspond to a progenitor cluster in the process of reaching complete dissolution \citep{balbinot18}. \citet{pricewhelan18} also notes that they could not rule out the scenario where the GD-1 progenitor dissolved long ago and the progenitor remnant exists in the form of the deep under-density seen in their density map at $\phi_1 = -20^{\circ}$. Alternatively, \citet{deboer18} argues that the progenitor remnant is located at $\phi_1 = -45^{\circ}$ due to their detection of an under density along the stream track surrounded by over-densities on either side using CHFT/Megacam photometry.

In this study, we perform a large suite of direct N-body simulations of star clusters with the same orbit as the GD-1 stream in a Milky Way-like potential that range in initial mass, size, and age to determine the most likely position for the (possibly dissolved) progenitor. From the collection of models, we find the sets of initial conditions that best reproduce a stellar stream with the same location, length and width as the observed GD-1 stream. From our best-fit simulations, we can then identify where the GD-1 progenitor or its remnant is currently located along the stream and what observational signature it will have. In Section \ref{s_method} we describe our analysis of the \gaia\ DR2 data on GD-1 to extract a new orbital fit to GD-1 that forms the basis of our simulations and to determine the density profile along the stream that we use to constrain the simulations. In Section \ref{s_nbody}, we give the details of our star cluster simulations. In Section \ref{s_results} we compare the length and position of our model streams, as well as the location and width of individual gaps along the stream, to the GD-1 stream itself. Finally, in Section \ref{s_discussion} we constrain the location of GD-1's progenitor based on our simulations and summarize our findings.

\section{GD-1 data}\label{s_method}

To search for the progenitor of the GD-1 stream, we generate a suite of direct $N$-body star cluster simulations with a range of initial masses, sizes, and ages in an attempt to reproduce a GD-1-like stream. Comparing the properties of model streams with different progenitor locations to properties of the observed GD-1 stream will allow for constraints to be placed on the exact location of GD-1's progenitor. However before we can simulate the evolution of GD-1's progenitor, we must first know its orbital history. Second, we need to have an observational dataset of stream stars to compare our simulations with. In the following, we describe how we determine our assumed orbit for the GD-1 stream and our method for extracting GD-1 stars from the \gaia\ DR2 data.

\subsection{The orbit of GD-1} \label{s_orbit}

Two recent estimates of the stream's orbit come from a combination of Sloan Digital Sky Survey (SDSS) and USNO-B data \citep{koposov10,bovy16_mwmap}. We combine the data from these previous estimates with the positions and proper motions of GD-1 stream stars in the \gaia\ DR2 catalogue \citep{Gaia16a,Gaia18} to yield a precise estimate of GD-1's orbit.

To fit an orbit to the GD-1 stream, we make use of the orbit fitting routine in \texttt{galpy} \footnote{http://github.com/jobovy/galpy} \citep{bovy15}, which fits a single orbit to a collection of stellar positions and velocities (and their associated errors) for a given potential using a marginalized $\chi^2$ method. For the purposes of this study, the potential of the Milky Way is taken to be the \texttt{MWPotential2014} model from \citet{bovy15}. The bulk of the stars used to fit an orbit to GD-1 come from the \gaia\ DR2 analysis by \citet{pricewhelan18}, which provides a catalogue of likely GD-1 members. We take the positions and proper motions of stars along the GD-1 track using the combination of their proper motion, photometric, and stream-track cuts. However for these stars, the uncertainty in their distances and line-of-sight velocities in \gaia\ DR2 (for those stars that have radial velocity measurements) is quite large. Therefore we only consider the positions and proper motions of these stars when estimating GD-1's orbit. To compliment the \gaia\ dataset, we also make use of six estimates of the distance to stars along the stream and twenty three measurements of the line-of-sight velocity of stars along the stream from \citet{koposov10}.

We find the best six dimensional orbital parameters near the middle of the observed stream to be given by 
\begin{align}
    \mathrm{RA} & = \phantom{-}148.9363998668805^{\circ}\,\\
    \mathrm{Dec} & = 36.15980426805254^{\circ}\,\\
    \mathrm{Distance} & = \phantom{-}7.555339165941959\,\mathrm{kpc}\,\\
    \mu_{\mathrm{RA}}\,cos\, \mathrm{Dec} & = -5.332929760383195\,\mathrm{mas\,yr}^{-1}\,\\
    \mu_{\mathrm{Dec}} & = -12.198914465325117\,\mathrm{mas\,yr}^{-1}\,\\
    v_{\mathrm{los}} & = \phantom{-}6.944006091929623\,\mathrm{km\,s}^{-1}\,.
\end{align}

\begin{figure}
    \includegraphics[width=0.48\textwidth]{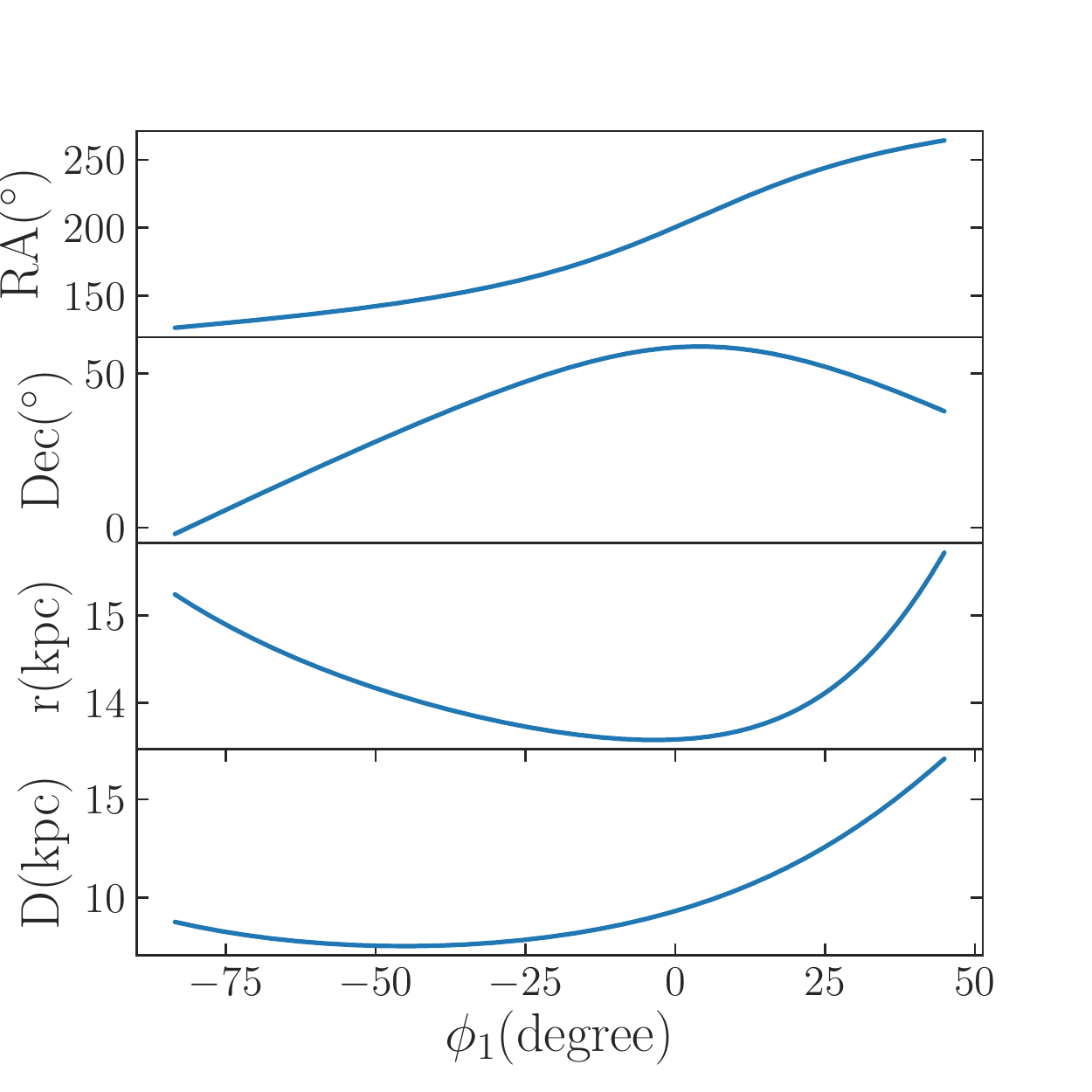}
    \caption{The last 100 Myr of GD1's orbit in terms of $\mathrm{RA}$, $\mathrm{Dec}$, distance from the sun ($\mathrm{d}$) and Galactocentric distance $\mathrm{r}$ as functions of $\phi_1$.}
    \label{fig:orbit}
\end{figure}

These parameters lead to an orbit for GD-1 that has a pericenter radius, apocenter radius and orbital eccentricity of approximately of 13.5 kpc, 23.5 kpc, and 0.27 respectively. Given these values, it is possible to integrate the stream's orbit backwards in time and generate model streams of different ages. For illustrative purposes, we show in Figure \ref{fig:orbit} the last 100 Myr of GD1's orbit. It is interesting to note that the stream is currently undergoing a perigalactic pass, with the leading edge of the stream having passed through pericentre approximately 50 Myr ago. In fact, stars located at $\phi_1 \sim -3.1$ are currently at pericenter. The previous perigalactic pass occurred 487 Myr ago.

We emphasize that we use this orbit fit only to get the past orbit of GD-1 approximately correct. In detail, the observed stream location does not follow a single orbit exactly \citep[e.g.,][]{Eyre11a,Sanders13a,Bovy14a}, which can have a large impact when constraining the Galactic potential using streams, but in practice for GD-1 the difference between the orbit and the stream is small \citep{Sanders13a,bovy16_mwmap}. The past orbit also depends on knowing the Galactic potential in the region that the GD-1 stream passes through and changes to the Galactic potential will lead to slightly different past orbits. Because we only attempt to roughly match overall characteristics of the GD-1 stream below, small changes in the past orbit do not have a significant impact on our results.

\subsection{GD-1 stream determination using \emph{Gaia} DR2}\label{s_gaia}

\begin{figure*}
	\centering
    \includegraphics[width=0.8\hsize]{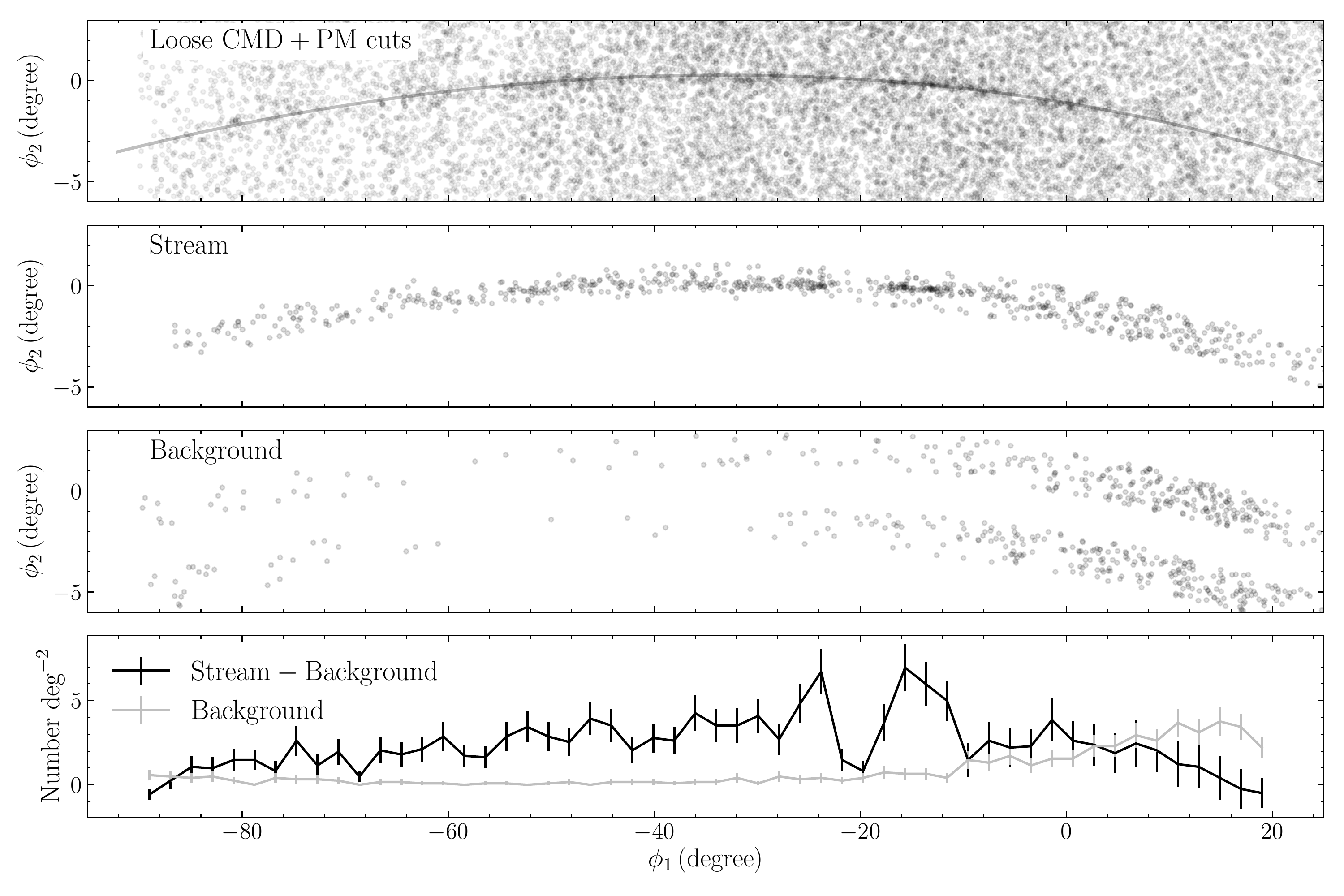}
    \caption{The GD-1 stream in \gaia\ DR2. The top panel shows stars selected using a loose colour-magnitude and proper-motion selection; these cuts clearly reveal the stream and the orbit fit from Section \ref{s_orbit} is shown for comparison. Using a sliding set of cuts in colour-magnitude and proper motion, we select a clean sample of GD-1 candidate stars, shown in the second panel. An estimate of the background is obtained using the same cuts applied away from the stream, this is displayed in the third panel. The resulting background-subtracted density profile as a function of $\phi_1$ for GD-1 is given in the bottom panel.}
    \label{fig:GD1stars}
\end{figure*}

Given an orbital path for the GD-1 stream, it is now possible to generate a refined dataset of GD-1 stream stars based on their positions, proper motions, colours and magnitudes that we use to determine the length, extent, and density of the GD-1 stream. We select stars by first querying the \gaia\ DR2 Archive for all stars with parallax $\varpi < 1\,\mathrm{mas}$, $-90^\circ < \phi_1 < 30^\circ$, $-6.5^\circ < \phi_2 < 3.5^\circ$, $-16\,\mathrm{mas\,yr}^{-1} < \mu_{\phi_1}\,\cos \phi_2 < -6\,\mathrm{mas\,yr}^{-1}$, and $-10\,\mathrm{mas\,yr}^{-1} < \mu_{\phi_2} < 5\,\mathrm{mas\,yr}^{-1}$, because these cuts encompass the observed GD-1 stream. We further require all stars to be matched to the Pan-STARRS catalogue \citep{chambers16}. We make these cuts by rotating all stars with $\varpi < 1\,\mathrm{mas}$ to the $(\phi_1,\phi_2$ coordinate system and performing the cuts directly in SQL with the query given in the Appendix. We then use a colour-magnitude filter consisting of a PARSEC isochrone in the Pan-STARRS magnitude system \citep{Bressan12a} for a metallicity [M/H] = $-1.35$ and an age of 12 Gyr. Placing this colour-magnitude filter at a distance of 8 kpc and selecting stars within $\Delta g-r = \pm 0.05$ and $g > 18$ clearly reveals the stream; this is shown in the top panel of Figure \ref{fig:GD1stars}. 

We select GD-1 candidate stars by sliding the colour-magnitude filter along the distance as a function of $\phi_1$ coming from the GD-1 orbit fit in Section \ref{s_orbit} and selecting stars with $\Delta g-r = \pm 0.03\pm 0.05/3\times (g-18)$ with $18 < g < 21$ which are further within $\Delta \mu_{\phi_1}\,\cos\phi_2 = ^{+1.5}_{-1.0}\,\mathrm{mas\,yr}^{-1}$, $\Delta \mu_{\phi_2} = \pm1.0\,\mathrm{mas\,yr}^{-1}$, $\Delta \phi_1 = \pm 1^{\circ}$ from the orbit fit. The reason for the asymmetric $\Delta \mu_{\phi_1}$ cut is that the stream is slightly offset from the orbit fit in this coordinate. Stars selected using this cut are displayed in the second panel of Figure \ref{fig:GD1stars}. To estimate the background, we use the same cuts, except for the $\Delta \phi_1$ cut, where we instead select stars that are between 1 and 3 degrees away from the orbit fit on either side; these background stars are shown in the third panel of Figure 3. It is clear that for most of the region covered by the stream, the background is very low, with the background only becoming important at $\phi_1 \gtrsim 0^\circ$. 

We obtain the background-subtracted density as a function of $\phi_1$ for GD-1 by binning the stream candidates and background stars in $\Delta \phi_1 = 2^\circ$ bins and subtracting the density of the background in each bin. This density is shown in the bottom panel of Figure \ref{fig:GD1stars}. We also show the density of the background, which is very low at $\phi_1 \lesssim 0^\circ$, but becomes important at the trailing end of the stream. 

The key observational properties of the GD-1 stream that we are aiming to reproduce with the simulations are the location and length of the stream in the GD-1 coordinate system ($\phi_1$,$\phi_2$). Starting from the density of stream stars in the bottom panel of Figure \ref{fig:GD1stars}, we define bins with densities greater than or equal to ten percent of the maximum density to be part of the GD-1 stream, effectively excluding the low density ends of the dataset that are poorly constrained to be stream stars. We then define the location of the stream to be the bin with the maximum $\phi_1$ value ($\phi_{1,\mathrm{max}}$) and the length of the stream to be the distance between $\phi_{1,\mathrm{max}}$ and the bin with the minimum $\phi_1$ value ($\phi_{1,\mathrm{min}}$). This yields  $\phi_{1,\mathrm{max}} = 12.9^{\circ}$ and a length of $97.8^{\circ}$. Given our orbital fit to the GD-1 stream, this angular stream length corresponds to a physical length of approximately 13.9 kpc. It should be noted that excluding the density peaks on either side of the well defined gap at $\phi_1 = -20^\circ$ when calculating the maximum density only shifts the location of $\phi_{1,\mathrm{max}}$, and therefore our estimate of the streams length, by $2^{\circ}$.

Due to the higher background density at the trailing edge of the stream, a determination of $\phi_{1,\mathrm{max}}$ is more uncertain than $\phi_{1,\mathrm{min}}$. However we can use the fact that GD-1 is at pericenter and also in the process of turning around to perform a sanity check on our determination of $\phi_{1,\mathrm{max}}$. As illustrated in Figure \ref{fig:orbit}, due to the projection of GD-1's orbit on to the plane of the sky, stream stars with $\phi_1 < -44.8 ^{\circ}$ are currently moving away from us while stream stars with $\phi_1 > -44.8 ^{\circ}$ are moving towards us. This projection of GD-1's orbit means that stars in the trailing tail are moving radially towards us more than they are moving tangentially across the sky. Stars that are still approaching pericentre ($\phi_1 > -3.1$) are even more strongly affected by this projection effect, such that stars in the trailing edge of the stream have little tangential motion. As our models of long GD-1-like streams will confirm (see Section \ref{s_results}), a longer trailing stream would cause an increase in the number of stars at $12.9^{\circ}$ as the trailing tail approaches pericentre and the turnaround point, not a lengthening of the stream in the positive $\phi_1$ direction. This point is demonstrated in Figure \ref{fig:orbit} by the fact that for $\phi_1 > 13^{\circ}$, the rate at which the orbit moves across the plane of the sky is much smaller than the rate at which the orbit is approaching the sun. Since a building up of stars at positive values of $\phi_1$ is clearly not observed in Figure \ref{fig:GD1stars}, we are confident that our measurement of $\phi_{1,\mathrm{max}}$ is a robust estimate of the GD-1 stream's trailing edge.

\section{Direct N-body modeling} \label{s_nbody}

To simulate the evolution of progenitor star clusters that  produce a GD-1-like stream, we use a modified version of the direct $N$-body code \texttt{NBODY6} \citep{aarseth03} that allows for a power-law density spherical potential with an exponential cut-off to be used for the Galactic bulge and a NFW potential \citep{navarro96} for the Galactic halo. These modifications, along with a \citet{miyamoto75} potential for the Galactic disc, allow for \texttt{MWPotential2014} to be used as the background potential in our simulations.

Initial cluster conditions are setup using McLuster \citep{kupper11}, where we assume that the initial cluster is represented by a \citet{king66} model with $W_0$=2. Clusters are generated with either 12,000, 16,000, 20,000 or 24,000 stars and initial half-mass radii of 10 pc, 20 pc, and 30 pc. The masses of individual stars are drawn from a power-law initial mass function with a slope of -1.35 for masses between 0.1 and $50\,M_\odot$ and then evolved for 12 Gyr assuming a stellar metalicity of 0.001. Then, given our best fit orbit to the GD-1 stream, we assign model clusters with initial positions and velocities in the galaxy such that the progenitor cluster is located near the leading edge of the observed GD-1 stream after 1700 Myr, 2560 Myr, or 3400 Myr. Finally, we also consider 16,000 star models that evolve for 5125 Myr to explore how even longer lifetimes will affect the observed properties of model streams. Due to the fact that we only focus on stream ages less than 5125 Myr, our decision to evolve the masses of individual stars before running the simulation is acceptable since stellar evolution only affects the early evolution of star clusters.

\section{Results}\label{s_results}

To constrain the location of the GD-1 progenitor, we find for each simulation the location and length of the model stream that forms in the exact same way that we did for the observed stream with one caveat. Since several model streams still have a dense progenitor, we exclude bins within $2.5^{\circ}$ of the progenitors location when calculating the maximum density along the stream. This approach ensures that the relative density between the edges and the central portion of the stream are the same as in the observed data.

We find that only models where the progenitor cluster is able to evolve for 3400 Myr in the Galactic potential produce stellar streams that are of comparable length and width to the actual GD-1 stream by the time they reach the stream's current location. Younger clusters will in general have stream lengths that are too short, as stars have less time to escape the cluster and begin populating its tidal tails. Similarly, older clusters will in general produce stream lengths that are too long. However it is important to note that for a given cluster age, initially extended clusters start losing mass immediately and will have slightly older tails that are both longer and thicker than clusters with initially higher densities that must expand before stars can escape. Hence it is possible, given a high enough initial density, for older clusters to produce streams that are comparable to GD1. However the cluster would have to expand and reach a density that is comparable to the aged 3400 Myr model clusters considered here, at the appropriate age, to produce a GD-1-like stream. Hence the aged 3400 Myr model clusters end up representing the later evolution of older, high-density clusters anyway. The older model clusters (5125 Myr) that begin forming tails immediately also confirmed our statement that a longer stream results in stars building-up near $\phi_1 = 13^{\circ}$, as opposed to extending the stream in the positive $\phi_1$ direction, due to the projection of GD-1's orbit causing the trailing stream to be primarily moving towards us. Therefore, in what follows, we only consider models that evolve for 3400 Myr.

\begin{figure*}
    \includegraphics[width=\textwidth]{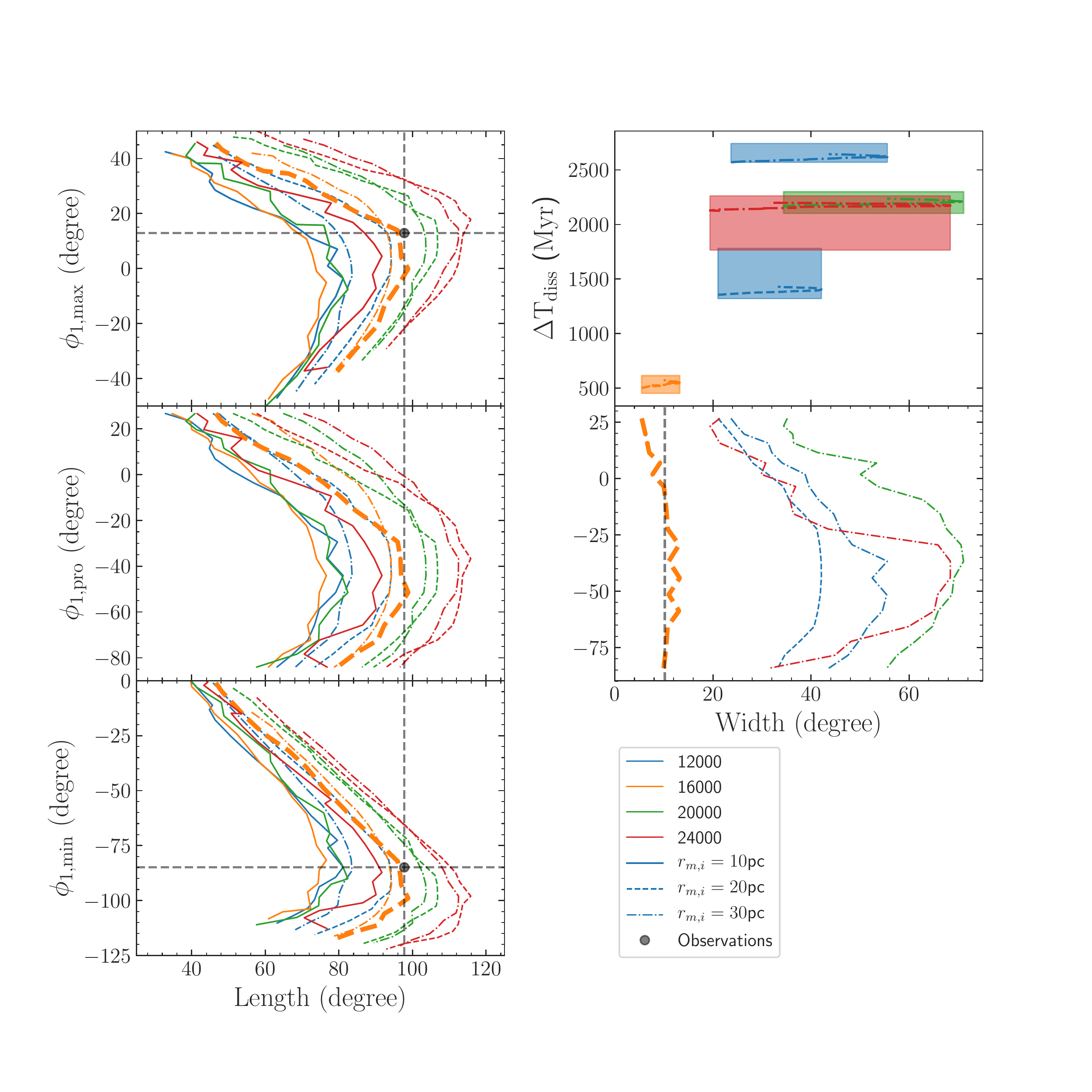}
    \caption{Location of GD-1's progenitor and the key observables. The left panels show the stream length as a function of $\phi_{1,\mathrm{max}}$ (top panel), $\phi_{1,\mathrm{pro}}$ (middle panel) and $\phi_{1,\mathrm{min}}$ (bottom panel) for model streams as they evolve from 3325 Myr to 3400 Myr of age. The initial size and number of stars in each simulation is noted in the legend and the observed length and $(\phi_{1,\mathrm{min}},\phi_{1,\mathrm{max}})$ values are indicated with grey points and lines. The right panels display the width of the gap that grows after the progenitor fully dissolves in our simulation as a function of time since dissolution $\Delta T_{\mathrm{diss}}$ (top panel) and $\phi_{1,\mathrm{pro}}$ (middle panel) for select models with clearly identifiable progenitor gaps. The width is compared to the width of the gap around $\phi_1 =-40^\circ$ in the \gaia\ data. The dissolution time is defined as the time where the number of bound stars first drops to below $1\,\%$ of the initial number of stars in the cluster. The shaded box around each line outlines the region where $0.5\,\%$ and $2\,\%$ of the initial number of stars are still bound. The GD-1 data favour a progenitor location in the range $-30^{\circ} < \phi_{1,\mathrm{pro}} < -45^{\circ}$; if the progenitor dissolved after its previous pericentric disc passage $\approx 500\,\mathrm{Myr}$ ago, the width of the gap around $\phi_1 =-40^\circ$ in the \gaia\ data is also naturally explained.
   \label{fig:6plot}}
\end{figure*}

The key observable parameters of the 3400 Myr old model streams are summarized in Figure \ref{fig:6plot}. The left panels illustrate the final 75 Myr of evolution of each stream's length as a function of $\phi_{1,\mathrm{max}}$ (top left panel), the location of the progenitor $\phi_{1,\mathrm{pro}}$ (middle left panel), and $\phi_{1,\mathrm{min}}$ (bottom left panel). This time range covers the progenitor's evolution across the observed length of the GD-1 stream. In general, we find that initially more compact clusters have shorter streams than extended clusters as they require time to expand before stars can escape and start populating the tidal tails. Lower-mass clusters also have shorter streams since they reach dissolution faster than high mass clusters and stars stop populating the stream. Comparing the distribution of model stream lengths and locations to the actual GD-1 stream properties (marked as either a grey point or dashed grey line in the left panels of Figure \ref{fig:6plot}), we see that the best-fit progenitor clusters appear to have initially consisted of between 12,000 and 16,000 stars and had initial half-mass radii between 20 pc and 30 pc. The widths of the best-fitting model streams are also consistent with the GD-1 stream, as the full width at half maximum along the stream track is approximately $0.5^{\circ}$ in each case. 

The model which best reproduces the GD-1 stream (marked as a thick yellow line in Figure \ref{fig:6plot}), initially consists of 16,000 stars and has an initial size of 20 pc. When the model progenitor reaches the location of the trailing end of the GD-1 stream, the model stream's length is initially short ($67.6^{\circ}$ or 7.5 kpc) due to it recently being compressed at pericenter (see \citealt{kaderali18} for a detailed explanation of this process). The stream is able to extend in length to $94.1^{\circ}$ (12.7 kpc) by the time the progenitor reaches $\phi_1 = -32\pm 1^{\circ}$ and the stream has the same $\phi_{1,\mathrm{min}}$ and $\phi_{1,\mathrm{max}}$ as GD-1. There is of course some uncertainty in this estimate, as small changes to the cluster's initial size, mass, and age can lead to the progenitor being at a slightly different location while still yielding a stream with the same location and length as GD-1. It is clear from the middle, left panel of Figure \ref{fig:6plot} that the model stream's length and location remains largely consistent with that of the observed GD-1 stream for $-30^{\circ} < \phi_{1,\mathrm{pro}} < -45^{\circ}$, thus our initial conclusion is that the GD-1 progenitor must be within this range. Its exact location being sensitive to measurements of $\phi_{1,\mathrm{min}}$ and $\phi_{1,\mathrm{max}}$ for the observed stream, which are hampered by low stellar densities at the stream's edges, the exact initial conditions of our models, and possible perturbations to the stream density. 

The range in $\phi_{1,\mathrm{pro}}$ values that reproduces both the length and position of the stream is consistent with the gap in the density at $\phi_1=-40^{\circ}$ (see Figure \ref{fig:GD1stars}) and the estimate from \citet{deboer18} based an under-density along the track at $\phi_1=-45^{\circ}$. Conversely, our simulations appear to rule out the progenitor being located at the $\phi_1=-20^{\circ}$ gap or an over-density at $\phi_1 = -13^{\circ}$ \citep{pricewhelan18}. For $-20^\circ < \phi_{1,\mathrm{pro}} < -10^{\circ}$, model streams extend from $\phi_1=~-70^{\circ}$ to $\phi_1=30^{\circ}$ with significant density at $\phi_1\approx 20^{\circ}$, which is inconsistent with \gaia\ observations. Hence it is only possible for the progenitor to be in this range if the trailing edge of the stream has been completely disrupted.

To further constrain the location of the GD-1 progenitor, we consider the density of stars along the stream as a function of time. In initially denser models that are high in mass, such that the cluster has yet to fully dissolve, there is a clear over-density at $\phi_{1,\mathrm{pro}}$ that is significantly larger than observed. Conversely, in less dense models where the cluster has recently dissolved, there is an under-density at the progenitor cluster's would-be location as escaped stars continue to move away from $\phi_{1,\mathrm{pro}}$. Therefore, the location and width of a density peak or gap associated with GD-1's progenitor will also be correlated with the location and length of the stream, with the gap width corresponding to how long ago the cluster dissolved. For very low-density clusters that have dissolved a long time ago, all evidence of the progenitor will have completely disappeared as the initial gap has widened to be of the same size scale as the length of the stream while the gap itself has become less deep due to stars that escape the cluster just before dissolution mixing with stars that escaped the host much earlier.

In the middle right panel of Figure \ref{fig:6plot} we plot the evolution of the width of the gap associated with the progenitor as a function of $\phi_{1,\mathrm{pro}}$ for models where the cluster has recently dissolved. We compare the models to the width of the gap at $\phi_1=-40^{\circ}$ in the GD-1 stream, which is $10.2^{\circ}$, because this is the clearest gap within our suggested progenitor location range of $-30^{\circ} < \phi_{1,\mathrm{pro}} < -45^{\circ}$. For comparison purposes, the time since cluster dissolution $\Delta T_{\mathrm{diss}}$ is illustrated in the top right panel. We define dissolution as the time where the number of bound stars first decreases to below $1\,\%$ the initial number of stars in the cluster. The shaded regions in the panel cover the area where dissolution is defined as the first time the number of bound stars drops below $2\,\%$ and $0.5\,\%$ the initial number of stars in the cluster.

The right panels of Figure \ref{fig:6plot} demonstrate that our best fit model is also able to reproduce the width of the $\phi_1=-40^{\circ}$ gap when $\phi_{1,\mathrm{pro}}=-40^{\circ}$. The gap width corresponds to the cluster reaching dissolution approximately 500 Myr ago. Given the time it took for stars to initially begin migrating beyond the clusters tidal radius, the length of time over which the best fit model stream forms is approximately 2600 Myr. Other models that have comparable stream lengths and locations all have large gap widths at the progenitors location due to the clusters reaching dissolution over 1 Gyr ago. The evolution of gap width with time is sensitive to the cluster's dissolution time, as additional simulations of model clusters that are either slightly denser or of comparable density but higher in mass than our best fit model all still have clearly identifiable over-densities at $\phi_{1,\mathrm{pro}}$. Additional simulations of slightly less dense clusters or clusters that are of comparable density but lower in mass reach dissolution even quicker than our best fit model and already have gap widths larger than any gaps in the GD-1 stream by the time they reach GD-1's location. Therefore, while our models suggest the gap at $\phi_{1,\mathrm{pro}}=-40^{\circ}$ is the most likely position of the GD-1 progenitor remnant and that it dissolved 500 Myr ago, we cannot exclude the scenario that the progenitor is more generally between $-30^{\circ} < \phi_{1,\mathrm{pro}} < -45^{\circ}$ and that it reached dissolution so long ago that any signature of its exact location along the stream has been erased over time. 

\section{Discussion and Summary}\label{s_discussion}
We have generated a large suite of direct N-body star-cluster simulations with the purpose of reproducing a stellar stream with the same observed properties of the well-known GD-1 stream. The complete suite of simulations includes models that begin with between 12,000 and 24,000 stars, have initial sizes ranging from 10 pc to 30 pc, and ages between 1700 Myr and 5125 Myr.  The models are compared to properties of the GD-1 stream extracted from the \gaia\ DR2 data as described in Section \ref{s_method}.

\begin{figure}
    \includegraphics[width=0.48\textwidth]{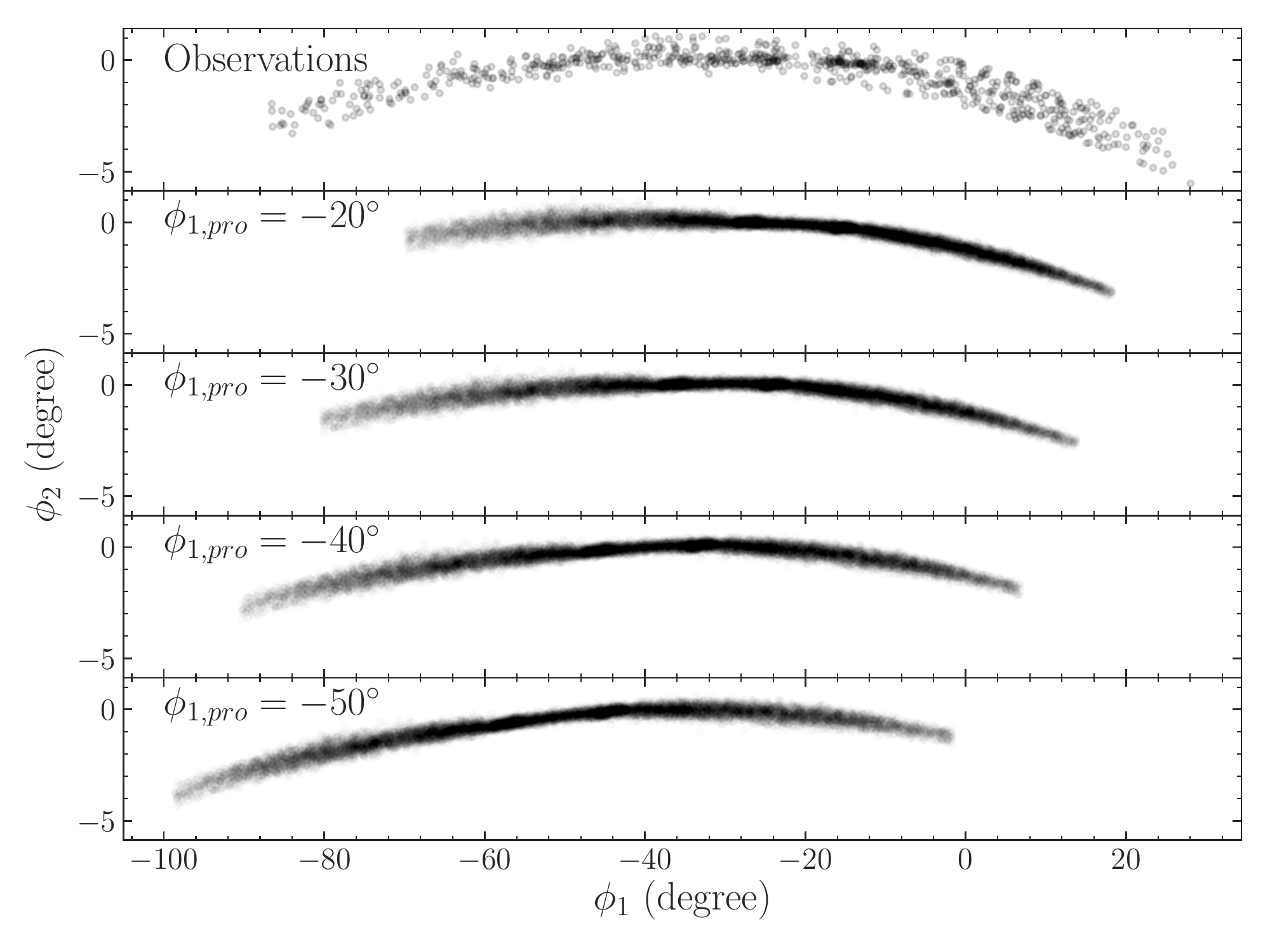}
    \caption{Observed location of the GD-1 stream in $\phi_1$ versus $\phi_2$ (top panel) compared to the stream in the best-fit simulation at various progenitor locations (lower panels). The model streams are similar to the observed stream for $\phi_{1,\mathrm{pro}}\approx -30^\circ$ and $\phi_{1,\mathrm{pro}} \approx -40^\circ$.
   \label{fig:phi1phi2_best}}
\end{figure}

The subset of models that yield stream locations and lengths comparable to GD-1 all start disrupting approximately 3400 Myr ago. Younger models yield stream lengths that are either too short. while older models would produce streams that are too long. The model that best reproduce a GD-1-like stream initially consisted of 16,000 stars and has an initial half-mass radius of 20 pc. As illustrated in Figure \ref{fig:phi1phi2_best}, after 3400 Myr the best-fit model stream reaches the same length and location as GD-1 when the progenitor is near $\phi_1 = -30^{\circ}$ ($\mathrm{RA}$, $\mathrm{Dec}$ = $156.7^{\circ}$, $44.1^{\circ}$). Given the fact that the model stream's length stays nearly constant as the progenitor moves along the stream's track and the uncertainty associated with measuring the exact end points of the GD-1 stream, we constrain the progenitors location to being between $-30^\circ < \phi_{1,\mathrm{pro}} < -45^{\circ}$. 

\begin{figure}
    \includegraphics[width=0.48\textwidth]{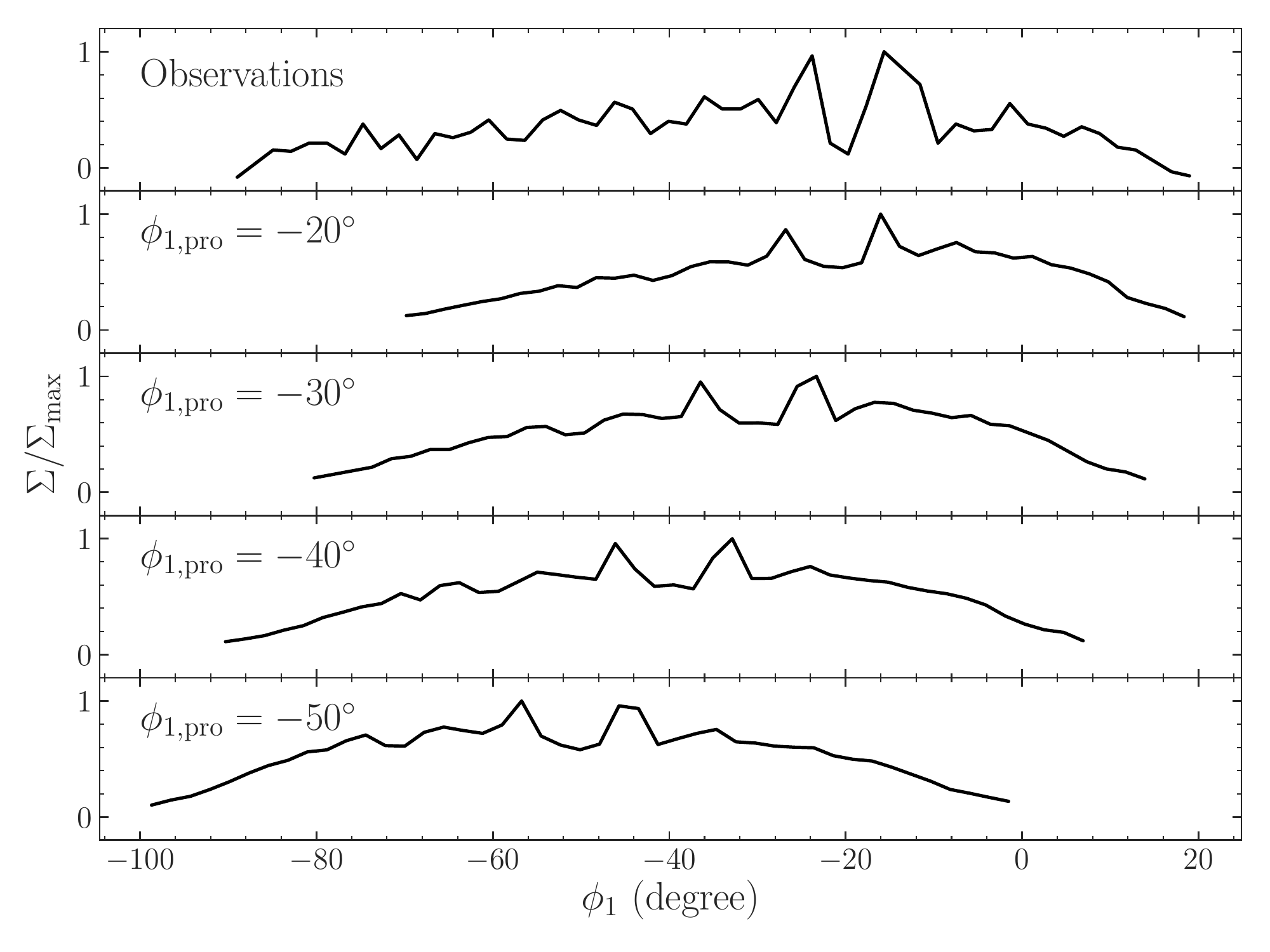}
    \caption{Surface density as a function of $\phi_1$ for the observed GD-1 stream (top panel) and for stars in the best-fit simulation at various progenitor locations (lower panels). If the progenitor has recently fully dissolved at $\phi_{1,\mathrm{pro}} \approx -40^\circ$, the dissolution leaves a gap that is simmilar to the observed gap at $\phi_1\approx -40^\circ$.
   \label{fig:phi1phi2_hist}}
\end{figure}

When taking into consideration the properties of the stream at the progenitor's location in our models, we are left with two different scenarios for the history of the progenitor cluster. The first scenario is that the progenitor cluster disrupted only recently, about 500 to 600 Myr ago, such that a small gap has formed along the stream where the progenitor would be; this gap forms as recently escaped stars continue to move away from the progenitor remnant with slightly higher and lower velocities than their host cluster had. In this scenario, the most likely location of the GD-1 progenitor is then $\phi_{1,\mathrm{pro}} = -40^{\circ}$ ($\mathrm{RA}$, $\mathrm{Dec}$ = $146.5^{\circ}$, $37.5^{\circ}$) where there exists a clear gap in the stream. As seen in Figure \ref{fig:phi1phi2_hist}, the location and width of the model cluster's gap nicely align with the observed gap around $\phi_1 = -40^{\circ}$ when the progenitor is at $\phi_{1,\mathrm{pro}} = -40^{\circ}$. The predicted dissolution time of the progenitor is consistent with GD-1's previous perigalactic pass, at which time the cluster also passed through the disc at a Galactocentric distance of about 13.5 kpc. Hence the strong tidal field experienced at perigalacticon is responsible for fully disrupting the progenitor cluster.

In the second scenario, the cluster reached dissolution over 2.5 Gyr ago such that the gap associated with the progenitor has since spread out to the point that it blends in with the naturally occurring density fluctuations along the stream. In this scenario we cannot constrain the progenitor's location beyond being between  $-30^\circ < \phi_{1,\mathrm{pro}} < -45^{\circ}$, consistent with recent estimates by \citet{deboer18}, and the GD-1 progenitor would have reached full dissolution only a few orbital periods after tidal disruption started, indicating that the cluster must have been of very low mass or concentration. An alternative explanation would then be required for the gap at  $\phi_1=-40^{\circ}$, which \citet{bonaca18} argues is due to a recent dark matter sub-halo interaction. However, the fact that the width of the gap and its location along the stream are consistent with a disruption event during the cluster's previous perigalactic pass suggests that the gap could instead be the result of a perturbing substructure in the Galactic disc (e.g., a concentrated molecular cloud). 
In both scenarios the stream's length and width are consistent with the progenitor cluster being actively stripped of stars for between approximately 2000 and 3000 Myr. It is interesting to note that a similar age is associated with the Pal 5 stream \citep{erkal17,bovy17}, which has a visible progenitor that is however close to full dissolution as well \citep{Dehnen04a}. This similarity suggests that Pal 5 and GD-1's progenitor clusters are both accreted clusters that joined the Milky Way during the same merger event. Constraining the ages of additional stellar streams and tidal tails may further help piece together the Milky Way's minor merger history, beyond what can be inferred from the ages and metallicities of its globular cluster population \citep[e.g.][]{kruijssen18}.

The fact that our models offer an explanation for a major gap along the GD-1 stream that does not invoke dark matter substructure highlights the importance of understanding the evolution of a stream's progenitor before the stream can be used to constrain the properties of sub-halos in the Galaxy. Constraining GD-1's progenitor to being between $-30^\circ < \phi_{1,\mathrm{pro}} < -45^{\circ}$, and possibly even being located right at $\phi_{1,\mathrm{pro}} = -40^{\circ}$, represents an important step towards using other gaps and densities along the stream to study the Milky Way. In particular, we note that deep gap at $\phi_1 \approx -20^\circ$ cannot be explained as being associated with the progenitor's dissolution in our modeling and it therefore provides a good candidate for a gap stemming from a perturber. However with streams being sensitive to all manners of substructure, dark and luminous \citep{Amorisco16a,pearson17,banik18}, significant modelling of individual streams is required in order to use them as tools to study the Milky Way. 

\section*{Acknowledgements}

JW acknowledges financial support through a Natural Sciences and Engineering Research Council of Canada (NSERC) Postdoctoral Fellowship. JW and JB also acknowledge additional financial support from NSERC (funding reference number RGPIN-2015-05235) and an Ontario Early Researcher Award (ER16-12-061). This work was made possible in part by Lilly Endowment, Inc., through its support for the Indiana University Pervasive Technology Institute, and in part by the Indiana METACyt Initiative. The Indiana METACyt Initiative at IU is also supported in part by Lilly Endowment, Inc.

\bibliographystyle{mnras}
\bibliography{ref2}

\begin{thebibliography}{}
\makeatletter
\relax
\def\mn@urlcharsother{\let\do\@makeother \do\$\do\&\do\#\do\^\do\_\do\%\do\~}
\def\mn@doi{\begingroup\mn@urlcharsother \@ifnextchar [ {\mn@doi@}
  {\mn@doi@[]}}
\def\mn@doi@[#1]#2{\def\@tempa{#1}\ifx\@tempa\@empty \href
  {http://dx.doi.org/#2} {doi:#2}\else \href {http://dx.doi.org/#2} {#1}\fi
  \endgroup}
\def\mn@eprint#1#2{\mn@eprint@#1:#2::\@nil}
\def\mn@eprint@arXiv#1{\href {http://arxiv.org/abs/#1} {{\tt arXiv:#1}}}
\def\mn@eprint@dblp#1{\href {http://dblp.uni-trier.de/rec/bibtex/#1.xml}
  {dblp:#1}}
\def\mn@eprint@#1:#2:#3:#4\@nil{\def\@tempa {#1}\def\@tempb {#2}\def\@tempc
  {#3}\ifx \@tempc \@empty \let \@tempc \@tempb \let \@tempb \@tempa \fi \ifx
  \@tempb \@empty \def\@tempb {arXiv}\fi \@ifundefined
  {mn@eprint@\@tempb}{\@tempb:\@tempc}{\expandafter \expandafter \csname
  mn@eprint@\@tempb\endcsname \expandafter{\@tempc}}}

\bibitem[\protect\citeauthoryear{{Aarseth}}{{Aarseth}}{2003}]{aarseth03}
{Aarseth} S.~J.,  2003, {Gravitational N-Body Simulations}.
Cambridge, UK: Cambridge University Press

\bibitem[\protect\citeauthoryear{{Amorisco}, {G{\'o}mez}, {Vegetti}  \&
  {White}}{{Amorisco} et~al.}{2016}]{Amorisco16a}
{Amorisco} N.~C.,  {G{\'o}mez} F.~A.,  {Vegetti} S.,   {White} S.~D.~M.,  2016,
  \mn@doi [\mnras] {10.1093/mnrasl/slw148}, \href
  {http://adsabs.harvard.edu/abs/2016MNRAS.463L..17A} {463, L17}

\bibitem[\protect\citeauthoryear{{Balbinot} \& {Gieles}}{{Balbinot} \&
  {Gieles}}{2018}]{balbinot18}
{Balbinot} E.,  {Gieles} M.,  2018, \mn@doi [\mnras] {10.1093/mnras/stx2708},
  \href {http://adsabs.harvard.edu/abs/2018MNRAS.474.2479B} {474, 2479}

\bibitem[\protect\citeauthoryear{{Banik} \& {Bovy}}{{Banik} \&
  {Bovy}}{2018}]{banik18}
{Banik} N.,  {Bovy} J.,  2018, preprint, \href
  {http://adsabs.harvard.edu/abs/2018arXiv180909640B} {} (\mn@eprint {arXiv}
  {1809.09640})

\bibitem[\protect\citeauthoryear{{Banik}, {Bertone}, {Bovy}  \&
  {Bozorgnia}}{{Banik} et~al.}{2018}]{Banik18b}
{Banik} N.,  {Bertone} G.,  {Bovy} J.,   {Bozorgnia} N.,  2018, \mn@doi [\jcap]
  {10.1088/1475-7516/2018/07/061}, \href
  {http://adsabs.harvard.edu/abs/2018JCAP...07..061B} {7, 061}

\bibitem[\protect\citeauthoryear{{Bonaca}, {Hogg}, {Price-Whelan}  \&
  {Conroy}}{{Bonaca} et~al.}{2018}]{bonaca18}
{Bonaca} A.,  {Hogg} D.~W.,  {Price-Whelan} A.~M.,   {Conroy} C.,  2018,
  preprint, \href {http://adsabs.harvard.edu/abs/2018arXiv181103631B} {}
  (\mn@eprint {arXiv} {1811.03631})

\bibitem[\protect\citeauthoryear{{Bovy}}{{Bovy}}{2014}]{Bovy14a}
{Bovy} J.,  2014, \mn@doi [\apj] {10.1088/0004-637X/795/1/95}, \href
  {http://adsabs.harvard.edu/abs/2014ApJ...795...95B} {795, 95}

\bibitem[\protect\citeauthoryear{{Bovy}}{{Bovy}}{2015}]{bovy15}
{Bovy} J.,  2015, \mn@doi [\apjs] {10.1088/0067-0049/216/2/29}, \href
  {http://adsabs.harvard.edu/abs/2015ApJS..216...29B} {216, 29}

\bibitem[\protect\citeauthoryear{{Bovy}}{{Bovy}}{2016}]{bovy16}
{Bovy} J.,  2016, \mn@doi [Physical Review Letters]
  {10.1103/PhysRevLett.116.121301}, \href
  {http://adsabs.harvard.edu/abs/2016PhRvL.116l1301B} {116, 121301}

\bibitem[\protect\citeauthoryear{{Bovy}, {Bahmanyar}, {Fritz}  \&
  {Kallivayalil}}{{Bovy} et~al.}{2016}]{bovy16_mwmap}
{Bovy} J.,  {Bahmanyar} A.,  {Fritz} T.~K.,   {Kallivayalil} N.,  2016, \mn@doi
  [\apj] {10.3847/1538-4357/833/1/31}, \href
  {http://adsabs.harvard.edu/abs/2016ApJ...833...31B} {833, 31}

\bibitem[\protect\citeauthoryear{{Bovy}, {Erkal}  \& {Sanders}}{{Bovy}
  et~al.}{2017}]{bovy17}
{Bovy} J.,  {Erkal} D.,   {Sanders} J.~L.,  2017, \mn@doi [\mnras]
  {10.1093/mnras/stw3067}, \href
  {http://adsabs.harvard.edu/abs/2017MNRAS.466..628B} {466, 628}

\bibitem[\protect\citeauthoryear{{Bressan}, {Marigo}, {Girardi}, {Salasnich},
  {Dal Cero}, {Rubele}  \& {Nanni}}{{Bressan} et~al.}{2012}]{Bressan12a}
{Bressan} A.,  {Marigo} P.,  {Girardi} L.,  {Salasnich} B.,  {Dal Cero} C.,
  {Rubele} S.,   {Nanni} A.,  2012, \mn@doi [\mnras]
  {10.1111/j.1365-2966.2012.21948.x}, \href
  {http://adsabs.harvard.edu/abs/2012MNRAS.427..127B} {427, 127}

\bibitem[\protect\citeauthoryear{{Carlberg}}{{Carlberg}}{2009}]{carlberg09}
{Carlberg} R.~G.,  2009, \mn@doi [\apjl] {10.1088/0004-637X/705/2/L223}, \href
  {http://adsabs.harvard.edu/abs/2009ApJ...705L.223C} {705, L223}

\bibitem[\protect\citeauthoryear{{Carlberg}}{{Carlberg}}{2012}]{carlberg12}
{Carlberg} R.~G.,  2012, \mn@doi [\apj] {10.1088/0004-637X/748/1/20}, \href
  {http://adsabs.harvard.edu/abs/2012ApJ...748...20C} {748, 20}

\bibitem[\protect\citeauthoryear{{Carlberg}}{{Carlberg}}{2016}]{carlberg16}
{Carlberg} R.~G.,  2016, \mn@doi [\apj] {10.3847/0004-637X/820/1/45}, \href
  {http://adsabs.harvard.edu/abs/2016ApJ...820...45C} {820, 45}

\bibitem[\protect\citeauthoryear{{Carlberg}}{{Carlberg}}{2017}]{carlberg17}
{Carlberg} R.~G.,  2017, \mn@doi [\apj] {10.3847/1538-4357/aa6479}, \href
  {http://adsabs.harvard.edu/abs/2017ApJ...838...39C} {838, 39}

\bibitem[\protect\citeauthoryear{{Carlberg} \& {Grillmair}}{{Carlberg} \&
  {Grillmair}}{2013}]{carlberg13}
{Carlberg} R.~G.,  {Grillmair} C.~J.,  2013, \mn@doi [\apj]
  {10.1088/0004-637X/768/2/171}, \href
  {http://adsabs.harvard.edu/abs/2013ApJ...768..171C} {768, 171}

\bibitem[\protect\citeauthoryear{{Chambers} et~al.,}{{Chambers}
  et~al.}{2016}]{chambers16}
{Chambers} K.~C.,  et~al., 2016, preprint, \href
  {http://adsabs.harvard.edu/abs/2016arXiv161205560C} {} (\mn@eprint {arXiv}
  {1612.05560})

\bibitem[\protect\citeauthoryear{{Dehnen}, {Odenkirchen}, {Grebel}  \&
  {Rix}}{{Dehnen} et~al.}{2004}]{Dehnen04a}
{Dehnen} W.,  {Odenkirchen} M.,  {Grebel} E.~K.,   {Rix} H.-W.,  2004, \mn@doi
  [\aj] {10.1086/383214}, \href
  {http://adsabs.harvard.edu/abs/2004AJ....127.2753D} {127, 2753}

\bibitem[\protect\citeauthoryear{{Diemand}, {Kuhlen}, {Madau}, {Zemp}, {Moore},
  {Potter}  \& {Stadel}}{{Diemand} et~al.}{2008}]{diemand08}
{Diemand} J.,  {Kuhlen} M.,  {Madau} P.,  {Zemp} M.,  {Moore} B.,  {Potter} D.,
    {Stadel} J.,  2008, \mn@doi [\nat] {10.1038/nature07153}, \href
  {http://adsabs.harvard.edu/abs/2008Natur.454..735D} {454, 735}

\bibitem[\protect\citeauthoryear{{Erkal} \& {Belokurov}}{{Erkal} \&
  {Belokurov}}{2015a}]{erkal15a}
{Erkal} D.,  {Belokurov} V.,  2015a, \mn@doi [\mnras] {10.1093/mnras/stv655},
  \href {http://adsabs.harvard.edu/abs/2015MNRAS.450.1136E} {450, 1136}

\bibitem[\protect\citeauthoryear{{Erkal} \& {Belokurov}}{{Erkal} \&
  {Belokurov}}{2015b}]{erkal15b}
{Erkal} D.,  {Belokurov} V.,  2015b, \mn@doi [\mnras] {10.1093/mnras/stv2122},
  \href {http://adsabs.harvard.edu/abs/2015MNRAS.454.3542E} {454, 3542}

\bibitem[\protect\citeauthoryear{{Erkal}, {Belokurov}, {Bovy}  \&
  {Sanders}}{{Erkal} et~al.}{2016}]{erkal16}
{Erkal} D.,  {Belokurov} V.,  {Bovy} J.,   {Sanders} J.~L.,  2016, \mn@doi
  [\mnras] {10.1093/mnras/stw1957}, \href
  {http://adsabs.harvard.edu/abs/2016MNRAS.463..102E} {463, 102}

\bibitem[\protect\citeauthoryear{{Erkal}, {Koposov}  \& {Belokurov}}{{Erkal}
  et~al.}{2017}]{erkal17}
{Erkal} D.,  {Koposov} S.~E.,   {Belokurov} V.,  2017, \mn@doi [\mnras]
  {10.1093/mnras/stx1208}, \href
  {http://adsabs.harvard.edu/abs/2017MNRAS.470...60E} {470, 60}

\bibitem[\protect\citeauthoryear{{Eyre} \& {Binney}}{{Eyre} \&
  {Binney}}{2011}]{Eyre11a}
{Eyre} A.,  {Binney} J.,  2011, \mn@doi [\mnras]
  {10.1111/j.1365-2966.2011.18270.x}, \href
  {http://adsabs.harvard.edu/abs/2011MNRAS.413.1852E} {413, 1852}

\bibitem[\protect\citeauthoryear{{Gaia Collaboration} et~al.,}{{Gaia
  Collaboration} et~al.}{2016}]{Gaia16a}
{Gaia Collaboration} et~al., 2016, \mn@doi [\aap]
  {10.1051/0004-6361/201629272}, \href
  {http://adsabs.harvard.edu/abs/2016A%26A...595A...1G} {595, A1}

\bibitem[\protect\citeauthoryear{{Gaia Collaboration} et~al.,}{{Gaia
  Collaboration} et~al.}{2018}]{Gaia18}
{Gaia Collaboration} et~al., 2018, \mn@doi [\aap]
  {10.1051/0004-6361/201833051}, \href
  {http://adsabs.harvard.edu/abs/2018A%26A...616A...1G} {616, A1}

\bibitem[\protect\citeauthoryear{{Grillmair} \& {Carlin}}{{Grillmair} \&
  {Carlin}}{2016}]{grillmair16}
{Grillmair} C.~J.,  {Carlin} J.~L.,  2016, in {Newberg} H.~J.,  {Carlin} J.~L.,
   eds,  Astrophysics and Space Science Library Vol. 420, Tidal Streams in the
  Local Group and Beyond. p.~87 (\mn@eprint {arXiv} {1603.08936}),
  \mn@doi{10.1007/978-3-319-19336-6_4}

\bibitem[\protect\citeauthoryear{{Grillmair} \& {Dionatos}}{{Grillmair} \&
  {Dionatos}}{2006}]{grillmair06}
{Grillmair} C.~J.,  {Dionatos} O.,  2006, \mn@doi [\apjl] {10.1086/505111},
  \href {http://adsabs.harvard.edu/abs/2006ApJ...643L..17G} {643, L17}

\bibitem[\protect\citeauthoryear{{Ibata}, {Lewis}, {Irwin}  \& {Quinn}}{{Ibata}
  et~al.}{2002}]{ibata02}
{Ibata} R.~A.,  {Lewis} G.~F.,  {Irwin} M.~J.,   {Quinn} T.,  2002, \mn@doi
  [\mnras] {10.1046/j.1365-8711.2002.05358.x}, \href
  {http://adsabs.harvard.edu/abs/2002MNRAS.332..915I} {332, 915}

\bibitem[\protect\citeauthoryear{{Johnston}, {Spergel}  \& {Haydn}}{{Johnston}
  et~al.}{2002}]{johnston02}
{Johnston} K.~V.,  {Spergel} D.~N.,   {Haydn} C.,  2002, \mn@doi [\apj]
  {10.1086/339791}, \href {http://adsabs.harvard.edu/abs/2002ApJ...570..656J}
  {570, 656}

\bibitem[\protect\citeauthoryear{{Kaderali}, {Hunt}, {Webb}, {Price-Jones}  \&
  {Carlberg}}{{Kaderali} et~al.}{2018}]{kaderali18}
{Kaderali} S.,  {Hunt} J.~A.~S.,  {Webb} J.~J.,  {Price-Jones} N.,   {Carlberg}
  R.,  2018, arXiv e-prints, \href
  {http://adsabs.harvard.edu/abs/2018arXiv180904108K} {}

\bibitem[\protect\citeauthoryear{{King}}{{King}}{1966}]{king66}
{King} I.~R.,  1966, \mn@doi [\aj] {10.1086/109857}, \href
  {http://adsabs.harvard.edu/abs/1966AJ.....71...64K} {71, 64}

\bibitem[\protect\citeauthoryear{{Koposov}, {Rix}  \& {Hogg}}{{Koposov}
  et~al.}{2010}]{koposov10}
{Koposov} S.~E.,  {Rix} H.-W.,   {Hogg} D.~W.,  2010, \mn@doi [\apj]
  {10.1088/0004-637X/712/1/260}, \href
  {http://adsabs.harvard.edu/abs/2010ApJ...712..260K} {712, 260}

\bibitem[\protect\citeauthoryear{{Kruijssen}, {Pfeffer}, {Reina-Campos},
  {Crain}  \& {Bastian}}{{Kruijssen} et~al.}{2018}]{kruijssen18}
{Kruijssen} J.~M.~D.,  {Pfeffer} J.~L.,  {Reina-Campos} M.,  {Crain} R.~A.,
  {Bastian} N.,  2018, \mn@doi [\mnras] {10.1093/mnras/sty1609}, \href
  {http://adsabs.harvard.edu/abs/2018MNRAS.tmp.1537K} {}

\bibitem[\protect\citeauthoryear{{K{\"u}pper}, {Maschberger}, {Kroupa}  \&
  {Baumgardt}}{{K{\"u}pper} et~al.}{2011}]{kupper11}
{K{\"u}pper} A.~H.~W.,  {Maschberger} T.,  {Kroupa} P.,   {Baumgardt} H.,
  2011, \mn@doi [\mnras] {10.1111/j.1365-2966.2011.19412.x}, \href
  {http://adsabs.harvard.edu/abs/2011MNRAS.417.2300K} {417, 2300}

\bibitem[\protect\citeauthoryear{{Miyamoto} \& {Nagai}}{{Miyamoto} \&
  {Nagai}}{1975}]{miyamoto75}
{Miyamoto} M.,  {Nagai} R.,  1975, \pasj, \href
  {http://adsabs.harvard.edu/abs/1975PASJ...27..533M} {27, 533}

\bibitem[\protect\citeauthoryear{{Navarro}, {Frenk}  \& {White}}{{Navarro}
  et~al.}{1996}]{navarro96}
{Navarro} J.~F.,  {Frenk} C.~S.,   {White} S.~D.~M.,  1996, \mn@doi [\apj]
  {10.1086/177173}, \href {http://adsabs.harvard.edu/abs/1996ApJ...462..563N}
  {462, 563}

\bibitem[\protect\citeauthoryear{{Pearson}, {Price-Whelan}  \&
  {Johnston}}{{Pearson} et~al.}{2017}]{pearson17}
{Pearson} S.,  {Price-Whelan} A.~M.,   {Johnston} K.~V.,  2017, \mn@doi [Nature
  Astronomy] {10.1038/s41550-017-0220-3}, \href
  {http://adsabs.harvard.edu/abs/2017NatAs...1..633P} {1, 633}

\bibitem[\protect\citeauthoryear{{Price-Whelan} \& {Bonaca}}{{Price-Whelan} \&
  {Bonaca}}{2018}]{pricewhelan18}
{Price-Whelan} A.~M.,  {Bonaca} A.,  2018, \mn@doi [\apjl]
  {10.3847/2041-8213/aad7b5}, \href
  {http://adsabs.harvard.edu/abs/2018ApJ...863L..20P} {863, L20}

\bibitem[\protect\citeauthoryear{{Sanders} \& {Binney}}{{Sanders} \&
  {Binney}}{2013}]{Sanders13a}
{Sanders} J.~L.,  {Binney} J.,  2013, \mn@doi [\mnras] {10.1093/mnras/stt806},
  \href {http://adsabs.harvard.edu/abs/2013MNRAS.433.1813S} {433, 1813}

\bibitem[\protect\citeauthoryear{{Sanders}, {Bovy}  \& {Erkal}}{{Sanders}
  et~al.}{2016}]{sanders16}
{Sanders} J.~L.,  {Bovy} J.,   {Erkal} D.,  2016, \mn@doi [\mnras]
  {10.1093/mnras/stw232}, \href
  {http://adsabs.harvard.edu/abs/2016MNRAS.457.3817S} {457, 3817}

\bibitem[\protect\citeauthoryear{{Springel} et~al.,}{{Springel}
  et~al.}{2008}]{springel08}
{Springel} V.,  et~al., 2008, \mn@doi [\mnras]
  {10.1111/j.1365-2966.2008.14066.x}, \href
  {http://adsabs.harvard.edu/abs/2008MNRAS.391.1685S} {391, 1685}

\bibitem[\protect\citeauthoryear{{Yoon}, {Johnston}  \& {Hogg}}{{Yoon}
  et~al.}{2011}]{yoon11}
{Yoon} J.~H.,  {Johnston} K.~V.,   {Hogg} D.~W.,  2011, \mn@doi [\apj]
  {10.1088/0004-637X/731/1/58}, \href
  {http://adsabs.harvard.edu/abs/2011ApJ...731...58Y} {731, 58}

\bibitem[\protect\citeauthoryear{{de Boer}, {Belokurov}, {Koposov},
  {Ferrarese}, {Erkal}, {C{\^o}t{\'e}}  \& {Navarro}}{{de Boer}
  et~al.}{2018}]{deboer18}
{de Boer} T.~J.~L.,  {Belokurov} V.,  {Koposov} S.~E.,  {Ferrarese} L.,
  {Erkal} D.,  {C{\^o}t{\'e}} P.,   {Navarro} J.~F.,  2018, \mn@doi [\mnras]
  {10.1093/mnras/sty677}, \href
  {http://adsabs.harvard.edu/abs/2018MNRAS.477.1893D} {477, 1893}

\makeatother
\end{thebibliography}

\bsp

\appendix

\clearpage

\section{\emph{Gaia} DR2 GD-1 query}

The basic query used in the \emph{Gaia} DR2 selection of likely GD-1 members discussed in Section \ref{s_gaia} is listed below for reference. Note that this query does not currently run on the actual Gaia Archive, because it times out, and to run it we have therefore run it on a local copy of the Gaia database.

\begin{lstlisting}[numbers=left]
SELECT 
    gaia.source_id,gaia.ra,gaia.dec,
    gaia.pmra,gaia.pmdec,
    gaia.parallax, gaia.parallax_error,
    gaia.phi1_fromsin,gaia.pmphi1,gaia.pmphi2,gaia.phi2,
    panstarrs1.g_mean_psf_mag as g, 
    panstarrs1.r_mean_psf_mag as r, 
    panstarrs1.i_mean_psf_mag as i, 
    panstarrs1.z_mean_psf_mag as z,
    panstarrs1.g_mean_psf_mag_error as gerr,
    panstarrs1.r_mean_psf_mag_error as rerr,
    panstarrs1.i_mean_psf_mag_error as ierr,
    panstarrs1.z_mean_psf_mag_error as zerr
FROM 
    (SELECT 
        source_id,ra,dec,pmra,pmdec,parallax,parallax_error,
        cosphi1cosphi2,sinphi2,asin(sinphi2) as phi2,
        asin(sinphi1cosphi2/cos(asin(sinphi2))) as phi1_fromsin,
        ( c1*pmra+c2*pmdec)/cos(asin(sinphi2)) as pmphi1, 
        (-c2*pmra+c1*pmdec)/cos(asin(sinphi2)) as pmphi2
    FROM 
        (SELECT                                 source_id,ra,dec,pmra,pmdec,
            parallax,parallax_error,
            -0.4776303088*cos(radians(dec))*cos(radians(ra))
                -0.1738432154*cos(radians(dec))*sin(radians(ra))
                +0.8611897727*sin(radians(dec)) as cosphi1cosphi2,
            0.510844589*cos(radians(dec))*cos(radians(ra))
                -0.8524449229*cos(radians(dec))*sin(radians(ra))
                +0.111245042*sin(radians(dec)) as sinphi1cosphi2,
            0.7147776536*cos(radians(dec))*cos(radians(ra))
                +0.4930681392*cos(radians(dec))*sin(radians(ra))
                +0.4959603976*sin(radians(dec)) as sinphi2,
            0.4959604136355941*cos(radians(dec))
                -0.8683451319068993*sin(radians(dec))*cos(radians(ra-34.59869021803964)) as c1,
            0.8683451319068993*sin(radians(ra-34.59869021803964)) as c2 
        FROM gaiadr2.gaia_source
        WHERE parallax < 1.) tab
    WHERE 
        sinphi2 >= -1 and sinphi2 <= 1.
        and sinphi1cosphi2/cos(asin(sinphi2)) >= -1.
        and sinphi1cosphi2/cos(asin(sinphi2)) <= 1.) gaia
INNER JOIN gaiadr2.panstarrs1_best_neighbour as panstarrs1_match ON panstarrs1_match.source_id = gaia.source_id
INNER JOIN gaiadr2.panstarrs1_original_valid as panstarrs1 ON panstarrs1.obj_id = panstarrs1_match.original_ext_source_id
WHERE 
    phi2 < radians(3.5) and phi2 > radians(-6.5) 
    and phi1_fromsin > radians(-90.) 
    and phi1_fromsin < radians(30.)
    and (cosphi1cosphi2/cos(asin(sinphi2))) > 0
    and pmphi1 > -16. and pmphi1 < -6 
    and pmphi2 > -10. and pmphi2 < 5.;
\end{lstlisting}

The constants in this query starting at line 23 are those that describe the transformation between $(\mathrm{RA},\mathrm{Dec})$ and $(\phi_1,\phi_2)$. Those in lines 23 through 31 are those from the transformation matrix given in the Appendix of \citet{koposov10}. The constants in line 32 through 34 are those necessary to directly transform proper motions from $(\mathrm{RA},\mathrm{Dec})$ to $(\phi_1,\phi_2)$. They are the sine and minus the cosine of the declination and minus the right ascension of the north pole of the $(\phi_1,\phi_2)$ frame in lines 32/33 and similar in line 34; these are all straightforwardly calculated from the transformation matrix of \citet{koposov10}. The query returns 332,323 stars.

\label{lastpage}

\end{document}